\documentclass[prb,preprint,showpacs,preprintnumbers,amsmath,amssymb]{revtex4}

\usepackage{amssymb}
\usepackage{amsmath}
\usepackage{graphicx}

\begin{document}

\title{Photo-induced,  non-equilibrium spin and charge polarization in quantum rings}

\author{Zhen-Gang Zhu and Jamal Berakdar}
\address{Institut f\"{u}r Physik, Martin-Luther-Universit\"{a}t
Halle-Wittenberg, Nanotechnikum-Weinberg, Heinrich-Damerow-St. 4,
06120 Halle, Germany}

\begin{abstract}
We investigate the spin-dependent dynamical response of a quantum
ring with  a spin-orbit interaction upon the application of linearly
polarized, picosecond, asymmetric electromagnetic  pulses. The
oscillations of the generated dipole moment are sensitive to the
parity of the occupation number  in the ring and to the strength of
the spin-orbit coupling. It is shown how the associated emission
spectrum can be controlled via the pulse strength or a gate voltage.
In addition, we inspect how a static magnetic flux can modify the
non-equilibrium dynamics. In presence of the spin-orbit interaction
and for a paramagnetic ring, the applied pulse results in a
spin-split, non-equilibrium local charge density. The resulting
temporal spin polarization is directed perpendicular to the
light-pulse polarization axis and oscillates periodically with the
frequency of the spin-split charge density. The spin-averaged
non-equilibrium charge density possesses a left-right symmetry with
respect to the pulse polarization axis. The calculations presented
here are applicable to nano-meter rings fabricated in heterojuctions
of  III-V and II-VI semiconductors containing several hundreds
electrons.
\end{abstract}

 \maketitle

\section{Introduction}
Advances in nanotechnology opened the way for the synthesis of  artificial
nanostructures with sizes smaller than the phase coherence length of the
charge carriers \cite{heizel}. The electronic properties of these systems
are  dominated by quantum effects and interferences \cite{imry}.
Particulary interesting are ring structures which served as a paradigm
for the demonstration of various aspects of quantum mechanics \cite{imry}.
Currently available phase-coherent rings vary in a wide range in size
and particle density \cite{exp1,exp2,exp3,exp4,exp5,nitta99}.
 On the theoretical side, various features of the equilibrium
properties of quantum rings are well understood and documented
 \cite{imry}. Recently
 no-equilibrium dynamics triggered by
 external time-dependent electromagnetic fields has been the subject of research
 \cite{t1,t2,t3,t4,t5,t5a,matos05,matos1,matos2,t6,t7,andrey06}.
  In particular it has been shown that
 irradiations with picosecond, time-asymmetric, low-intensity light fields generates
charge polarization and charge currents in a qualitatively different manner
than  in the case of applied harmonic laser fields.
Currently,  asymmetric pulses are producible with
a duration from few hundreds   femtoseconds up to nanoseconds  \cite{hcp,seqhcp1,seqhcp2,seqhcp3,seqhcp4}.
 The optical cycle of the electric field of the asymmetric pulse consists of a short
 half cycle followed by a much longer and weaker half cycle of
an opposite polarity. Hence, under certain conditions, the external field acts
as a unipolar pulse and therefore it is referred to as an half-cycle pulse (HCP).

In this study we focus on quantum rings as those fabricated out of a
dimensional electron gas  formed between  heterojuctions of  III-V
and II-VI semiconductors. Spin-orbit interaction (SOI) is crucial in
these materials.
 The influence of the SOI on the equilibrium
properties of these rings have already been
studied\cite{meir,chaplik} (for more recent works we refer to
\cite{splett,frustaglia,molnar,foldi,sheng,meijer}). In this work,
we shall consider the spin-dependent \emph{non-equilibrium} dynamic
of the ring with SOI driven by HCP's and in the presence of a
magnetic flux, a problem which, to our knowledge, has not been
addressed so far.
 As shown below the applied pulse triggers an oscillating  charge
polarization with frequencies dependent on the number of particles,
the strength of SOI, the intensity of the light and the applied
static magnetic flux. The energy scale is set by field-free
eigenfrequency of the ring. Furthermore, it is shown that even
though the light does not couple directly to the spin (at the
intensities considered here) the presence of SOI leads to a temporal
spin-splitting in the non-equilibrium local charge density. The
resulting non-equilibrium, local spin polarization is perpendicular
to the light-pulse polarization axis. It oscillates with the same
frequency of the spin-split charge density and thus its time average
vanishes. Experimentally, the induced polarization is measurable by
detecting the associated radiation emission. In fact, the driven
rings can be utilized as a source for harmonic generation. As shown
below the power spectrum is, to some extent
 tunable by an external static field that controls the strength of the
Rashba SO coupling.

This work is organized as follows, at first we shall derive the
Hamiltonian of the ring with SOI coupled to the HCP's field and in
the presence of a static magnetic flux. In Sec.II. we discuss the
initial carriers' wave functions, energies with and without SOI. In Sec.III. we
consider the time-dependent dynamics for the system when applying
the pulse field. In section IV, we present our calculations for the
dipole moment of the ring. Detailed numerical calculations and
discussions are contained in Sec. V.

\section{Quantum rings with spin orbit interaction}

\subsection{Hamiltonian}
We shall consider a quantum ring (QR) with a spin orbit interaction
subjected to a time-dependent electromagnetic field. In a minimal
coupling scheme the single-particle Hamiltonian \cite{meijer} reads
\begin{equation}
H_{0}=\frac{\mathbf{\Pi}^{2}}{2m^{*}}+V(\mathbf{r})-e\Phi+\frac{\alpha_{R}}{\hbar}
(\hat{\sigma}\times\mathbf{\Pi})_{z}+
\mathbf{\mu}_{B}\mathbf{B}\cdot\hat{\sigma}, \label{h1}
\end{equation}
where $\mathbf{\Pi}=\mathbf{p}+e\mathbf{A}$, $\mathbf{p}$ is
momentum operator, $e$ is the charge of the carrier, and $\mathbf{A}$ is
the vector potential of the external  electromagnetic (EM) field.  The  term $V(\mathbf{r})$ in Eq. (\ref{h1}) is the
 potential  confining the particles to the QR; the third
term in Eq. (\ref{h1}) is the scalar potential  $\Phi$ of  the EM
field,  the fourth term  is the Rashba SOI with the coupling constant $\alpha_{R}$; the components of
$\hat{\sigma}$ are Pauli matrices. The last term in Eq. (\ref{h1})
is the Zeeman term describing  the coupling between the electrons' magnetic moment $\mu_B$ and the
magnetic component of the EM field.

The electric and magnetic fields $\mathbf{E}(\mathbf{r},t)=
-\nabla\Phi(\mathbf{r},t)-
\partial\mathbf{A}(\mathbf{r},t)/\partial t$, and $
\mathbf{B}(\mathbf{r},t)=\nabla\times\mathbf{A}(\mathbf{r},t)$ are
invariant under the local gauge transformations \cite{quantum
optics,qo2}
$\Phi'(\mathbf{r},t)=\Phi(\mathbf{r},t)-\frac{\partial\chi(\mathbf{r},t)}{\partial
t}$ and
$\mathbf{A}'(\mathbf{r},t)=\mathbf{A}(\mathbf{r},t)+\nabla\chi(\mathbf{r},t)$.
Introducing a unitary operator
$\hat{R}=\exp(-ie\chi(\mathbf{r},t)/\hbar)$ we find
\begin{eqnarray}
\hat{R}\frac{\alpha_{R}}{\hbar}(\hat{\sigma}\times\mathbf{\Pi})_{z}\hat{R}^{\dag}&=&
\frac{\alpha_{R}}{\hbar}[\hat{\sigma}\times(e^{-ie\chi/\hbar}\mathbf{\Pi}e^{ie\chi/\hbar})]_{z}
\notag \\
&=&\frac{\alpha_{R}}{\hbar}[\hat{\mathbf{\sigma}}\times\mathbf{\Pi}']_{z},\quad
\mathbf{\Pi}'=\mathbf{p}+e\mathbf{A}'. \notag
\end{eqnarray}
The transformed Hamiltonian reads
\begin{equation}
\hat{H}'=\frac{\mathbf{\Pi}'^{2}}{2m^{*}}+V(\mathbf{r})-e\Phi'+
\frac{\alpha_{R}}{\hbar}(\hat{\mathbf{\sigma}}\times\mathbf{\Pi}')_{z}+
\frac{i\mu_{B}}{\hbar}(\mathbf{\Pi}'\times\mathbf{A})\cdot\hat{\mathbf{\sigma}}.
\label{h2}
\end{equation}
Within the Coulomb (or radiation) gauge, i.e.  $\Phi=0$ and
$\nabla\cdot\mathbf{A}=0$  we obtain
\begin{eqnarray}
\hat{H}'&=&\frac{[\mathbf{p}+e(\mathbf{A}+\nabla\chi)]^{2}}{2m^{*}}+V(\mathbf{r})+e\frac{\partial\chi}{\partial
t}+\frac{\alpha_{R}}{\hbar}\left[\hat{\mathbf{\sigma}}\times(\mathbf{p}+e(\mathbf{A}+\nabla\chi))\right]_{z}
\notag\\
&&+
\frac{i\mu_{B}}{\hbar}[(\mathbf{p}+e(\mathbf{A}+\nabla\chi))\times\mathbf{A}]\cdot\hat{\mathbf{\sigma}}.
\label{ha2}
\end{eqnarray}
In what follows we employ a plane-wave vector potential $\mathbf{A}=\mathbf{A}_{0}e^{i(\mathbf{k}\cdot\mathbf{r}-\omega
 t)}+c.c.$ and a gauge function
 $\chi(\mathbf{r},t)=-\mathbf{A}(t)\cdot\mathbf{r}$, where $\mathbf{k}$ and $\omega$ are the wave vector and the frequency of the EM field.
   Furthermore we note that in our case the light
propagates perpendicular to the plane of the ring. The thickness $d$
of the ring will be on the order of nanometers. Thus, in the present
study  $1\gg \mathbf{k}\cdot\mathbf{r}$, $r\approx d$ and the dipole
approximation is justified, even though the radius $a$ of the ring could
be in the micrometer range. With $
\nabla\chi(\mathbf{r},t)=-\mathbf{A}(t)$, and
$\frac{\partial\chi(\mathbf{r},t)}{\partial
t}=-\mathbf{r}\cdot\frac{\partial\mathbf{A}}{\partial
t}=-\mathbf{r}\cdot\mathbf{E}(t)$ we find
\begin{equation}
\hat{H}'=\hat{H}_{SOI}+\hat{H}_{1}(t),\label{h3}
\end{equation}
where
\begin{equation}
\hat{H}_{\text{SOI}}=\frac{\mathbf{p}^{2}}{2m^{*}}+V(\mathbf{r})+\frac{\alpha_{R}}{\hbar}(\hat{\mathbf{\sigma}}\times\mathbf{p})_{z}
, \label{hsoi}
\end{equation}
is the Hamiltonian of a quantum ring with spin orbit interaction,
and
\begin{equation}
\hat{H}_{1}(t)=-e\mathbf{r}\cdot\mathbf{E}(t)+\mu_{B}\mathbf{B}(t)\cdot\hat{\mathbf{\sigma}}.
\label{h1t}
\end{equation}
Switching over to cylindrical coordinates and integrating out
the radial dependence, $\hat{H}_{\text{SOI}}$ attains the form
\cite{meijer,frustaglia,molnar,foldi,sheng}
\begin{eqnarray}
\hat{H}_{\text{SOI}}&=&\frac{\hbar^{2}}{2m^{*}a^{2}}(i\partial_{\varphi}+\frac{\phi}{\phi_{0}})^{2}-
\frac{\alpha_{R}}{a}(\sigma_{x}\cos\varphi+\sigma_{y}\sin\varphi)(i\partial_{\varphi}+\frac{\phi}{\phi_{0}})
\notag\\
&&-i\frac{\alpha_{R}}{2a}(\sigma_{y}\cos\varphi-\sigma_{x}\sin\varphi)+\frac{\hbar\omega_{B}}{2}\sigma_{z},
\label{hsoi1}
\end{eqnarray}
where $\partial_{\varphi}=\frac{\partial}{\partial\varphi}$,
$\phi_{0}=h/e$ is the unit of flux, $\phi=B\pi a^{2}$ is the
magnetic flux threading the ring, $a$ is the radius of the ring, and
$\omega_{B}=2\mu_{B}B/\hbar$. In Eq. (\ref{hsoi1}) we added a static
magnetic field in addition to the applied light field.

\section{Spin-dependent Charge Polarization induced by a single EM pulse}
\subsection{Ground-state wavefunctions and the spectrum of the ring}
The energy spectrum of a QR with SOI has been discussed in several
works \cite{frustaglia,molnar,foldi,sheng}. Consistent with Ref.
[\onlinecite{splett}] we find for the angular  single particle wave
functions $\Psi_{n}^{S}(\varphi)$
 (the index $T$ refers to the transpose)
\begin{equation}
\Psi_{n}^{S}(\varphi)=e^{i(n+1/2)\varphi}\nu^{S}(\gamma,\varphi),
\label{wavefunction}
\end{equation}
where $S$ and $n$ denote the spin and the integer angular quantum numbers and
\begin{equation}
\nu^{S}(\gamma,\varphi)=(a^{S}e^{-i\varphi/2},
b^{S}e^{i\varphi/2})^{T} \label{wavefnu}
\end{equation}
are spinors in the angle-dependent local frame and
\begin{eqnarray}
a^{\uparrow}&=&\cos(\gamma/2),
b^{\uparrow}=\sin(\gamma/2),\notag\\
a^{\downarrow}&=&-\sin(\gamma/2), b^{\downarrow}=\cos(\gamma/2).
\label{ab}
\end{eqnarray}
In absence of the static magnetic field, the angle $\gamma$ is given
by (cf. Fig. 1)
$$ \tan\gamma=-Q_{R}=-\omega_{R}/\omega_{0},\
\hbar\omega_{R}=2\alpha_{R}/a, \
\hbar\omega_{0}=\hbar^{2}/(m^{*}a^{2}).$$
%%%%%%%%%%%%%%%%%%%%%%%%%%%%%%%%%%%%%%%%%%%%%%%%%%%%%%%%%%
\begin{figure}[tbph]
\centering \includegraphics[width =12 cm, height=7 cm]{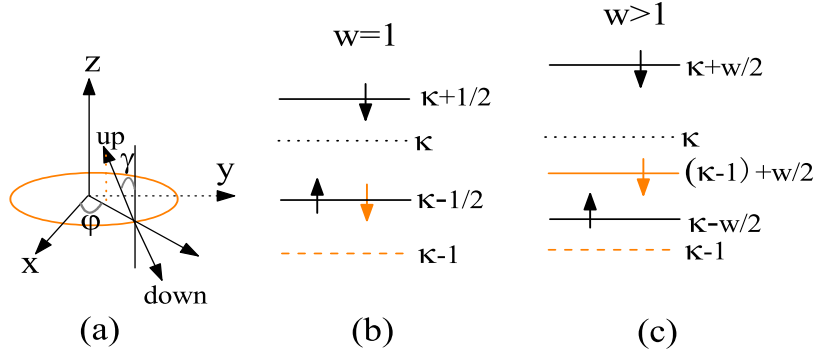}
\caption{(Color online) (a) Schematic illustration of the spin
orientation exhibited by the eigenstates of an ideal one-dimension
ring. (b) and (c) show  schematically  the energy spectrums for
 the \emph{total} angular momentum quantum number $\kappa$ at
different spin-orbit coupling strengths $w=\cos^{-1}\gamma$.}
\label{fig1}
\end{figure}
%%%%%%%%%%%%%%%%%%%%%%%%%%%%%%%%%%%%%%%%%%%%%%%%%%%%%%%%
The local spin orientations is inferred from the relations
\begin{equation}
S(\mathbf{r})_{\uparrow}=\frac{\hbar}{2}(\sin\gamma\cos\varphi
\hat{e}_{x}+\sin\gamma\sin\varphi
\hat{e}_{y}+\cos\gamma\hat{e}_{z}), \label{su}
\end{equation}
and
\begin{equation}
S(\mathbf{r})_{\downarrow}=\frac{\hbar}{2}[\sin(\pi-\gamma)\cos(\pi+\varphi)
\hat{e}_{x}+\sin(\pi-\gamma)\sin(\pi+\varphi)
\hat{e}_{y}+\cos(\pi-\gamma)\hat{e}_{z}]. \label{sdown}
\end{equation}
Thus $\gamma$ is the angle illustrated in Fig. 1(a). The limit
$Q_{R}\rightarrow\infty$, i.e. for a very strong SOI coupling,
$\gamma\rightarrow-\pi/2$ which corresponds to the plane of the
ring. The eigenenergies have the following structure
\begin{eqnarray}
E_{n}^{S}&=&\frac{\hbar\omega_{0}}{2} \left[(n-x_S)^{2}-\frac{Q_{R}^{2}}{4}\right],
\label{eigenenergy}\\
x_S&=&\frac{\phi}{\phi_{0}}-\frac{1-Sw}{2} \label{xs}\end{eqnarray}
where $w=\sqrt{1+Q_{R}^{2}}=1/\cos\gamma$, and $S=\pm1$ stand for up
and down spins. We emphasize that hereafter, the terms up and down
(labeled respectively $\uparrow$ and $\downarrow$) refer to
directions in the \emph{local} frame, as illustrated in Fig. 1.

\subsubsection{The weak SOI limit}
In the weak SOI limit, i.e. if ($\gamma\to 0$) and in absence of the
static external flux the eigenenergies attain the forms
$E_{n}^{\uparrow}=\frac{\hbar\omega_{0}}{2}n^{2}$ and
$E_{n}^{\downarrow}=\frac{\hbar\omega_{0}}{2}(n+1)^{2}$ which seems
inconsistent with the spin-degenerate ground-state formula
$E_{n}=\frac{\hbar\omega_{0}}{2}n^{2}$ (associated with
$\hat{H}_{0}=\frac{\hbar^{2}}{2m^{*}a^{2}}(i\partial_{\varphi})^{2}$).
The resolution of this apparent inconsistency is as follows.
$H_{\text{SOI}}$ commutes in a nontrivial way with $K=L_{z}+S_{z}$ which is
the \textit{z} component of the total angular momentum, and with
$S_{\gamma\varphi}=S_{x}\sin\gamma\cos\varphi+S_{y}\sin\gamma\sin\varphi+S_{z}\cos\gamma$
which is the spin component in the direction determined by the
angles $\gamma$ and $\varphi$. Also we can show that
$[K,S_{\gamma\varphi}]=0$. The simultaneous eigenfunctions of $H$,
$K$ and $S_{\gamma\varphi}$ are the function given by Eq.
(\ref{wavefunction}). To rotate the quantization axis of $S_{z}$ to
the direction $S_{\gamma\varphi}$, a SU(2) transformation is
necessary, i.e. $US_{z}U^{-1}=S_{\gamma\varphi}$, where $U$ is the
transformation matrix, $U_{11}=\cos\gamma/2\exp(-i\varphi/2)$,
$U_{12}=-\sin\gamma/2\exp(-i\varphi/2)$,
$U_{21}=\sin\gamma/2\exp(i\varphi/2)$, and
$U_{22}=\cos\gamma/2\exp(i\varphi/2)$.
Under the transformation of $U^{-1}$ the wave functions of SOI
become
$\Psi'=U^{-1}\Psi_{n}^{S}(\varphi)=e^{i(n+1/2)\varphi}\chi_{S}$,
where $\chi_{S}=(1,0)^{T}$ stands for spin-up state and $(0,1)^{T}$
for spin-down state. The exponential factor $n+1/2$ is the quantum
number for the $z$ component of the total angular momentum, referred
to as $K\Psi_{n}^{S}(\varphi)=\kappa\Psi_{n}^{S}(\varphi)$, where
$\kappa=n+1/2$. $\kappa$ is fixed in the process of the SU(2)
rotation. On the other hand, the wave function for $H_{0}$, the
Hamiltonian for a ring without SOI, is usually set as
$\Psi_{0}=e^{in\varphi}\chi_{S}$.\cite{matos1} Comparing $\Psi'$
and $\Psi_{0}$, it is clear that the exponential factors in $\Psi'$
and $\Psi_{0}$ are different, and $n$ in $\Psi_{0}$ is the quantum
number of orbital momentum. This difference manifests itself in the
limit $\gamma=0$ (i.e. SOI is zero, $H_{\text{SOI}}\to H_{0}$): the
wavefunctions $\Psi_{n}^{S}(\varphi)$ do not go over in $\Psi_{0}$
because they are the eigenfunctions for different sets of commuting
observables. The simultaneous set for $\Psi_{0}$ is $H$, $L_{z}$ and
$S_{z}$.

Despite this explanation we may still wonder what is the physical
effect of splitting of the energies on $n$ axis \cite{rashbajpc} in
the limit of vanishing SOI. We inspect therefore the persistent
charge current (CC) caused by SOI and take the limit of absence of
SOI in the last step. The  partial charge persistent  current due to
the state labeled by  $n$ and $S$ in the presence of  SOI and a static
external magnetic flux is
\begin{equation}
\mathbf{I}_{nS}=-\hat{\mathbf{e}}_{\varphi}I_{0}\left(n+\frac{\phi}{\phi_{0}}+\frac{1-Sw}{2}\right),
\label{pcc}
\end{equation}
where $I_{0}=2E_{0}a/\phi_{0}$ is the unit of current. When $\phi=0$,
the CC due to the particle in the $n$ level is $I_{nS}=-n$ for spin
up and $I_{nS}=-(n+1)$ in the limit of zero SOI (we ignored
$\hat{\mathbf{e}}_{\varphi}$ and scaled the current by $I_{0}$).
 The total CC is $I_{S}=\sum_{n}I_{nS}$. For a distribution of
up spins, the occupied states are $n=0,\pm1,\pm2,\cdots$; for spin-down particles, they are $n=-1,(0,-2 ),(1,-3 ),\cdots$; the numbers in the
same parenthesis ``(...)" indicate  states with the same energy.
Hence, as expected, we conclude that the total CCs for up and down
spins carriers are zero in the limit of vanishing external magnetic field and
SOI. Note, that the lowest occupied state for down spin is not at
$n=0$, but at $n=-1$. This means the velocity of the particle with
down spin at $n=-1$ is zero; in fact $n+1$ stands for the quantized
velocity for spin-down particles. Thus, shifting the energy spectrum in the down
spin channel to the right by 1 quantum number, the energy spectrums for
different spins become the same on the velocity axis which is
consistent with the physical picture \cite{rashbajpc}.

In presence of SOI the energy spectrum is rewritable as
\begin{equation}
E_{n}^{S}=\frac{\hbar\omega_{0}}{2}[(\kappa-Sw/2-\phi/\phi_{0})^{2}-Q_{R}^{2}/4]=
\frac{\hbar\omega_{0}}{2}[(\kappa-\frac 1 2 - x_S)^{2}-Q_{R}^{2}/4].
\label{e1}
\end{equation}
For $\phi=0$ the energies associated with a total angular momentum
quantum number $\kappa$ are spin-split; up spins have lower energy
than down spins  as illustrated in Fig. 1(b) and 1(c). For the case
of zero SOI, i.e. for $w=1$ (cf. Fig. 1(b)) electrons with spin-up
and  $\kappa$ are degenerate with electrons having spin-down and
$\kappa-1$ quantum numbers; the energy is given by $\kappa-1/2$,
meaning that the energy levels are spin-degenerate.

\subsection{Time-dependent wave functions and energies after  pulse irradiation}
As have been shown in detail
\cite{matos1,matos2,mizushima,berakdar}, upon applying at $t=0$ a
half-cycle pulse with a duration $\tau_d$, the time-dependent
electronic states of the ring develop as
\begin{equation}
\Psi_{n}^{S}(\varphi,t_0=\tau_d)=\Psi_{n}^{S}(\varphi,t=0)e^{i\alpha\cos\varphi}.
\label{matching}
\end{equation}
The parameter
\begin{equation}
\alpha=e\, a\, p/\hbar,\; p=-\int_{0}^{\tau_d} E(t) \rm{d}t
\label{alpha}
\end{equation}
is the action (in units of action) taken over by the carriers from
the pulse EM field. The range of validity of the the solution
(\ref{matching}) has been discussed in Refs.
[\onlinecite{matos1,matos2,berakdar}] (for a general discussion of
the properties of the time development operator we refer to the work
[\onlinecite{mizushima}]). The coherent state (\ref{matching}) is
not an eigenstate of the ring for $t>t_0$, in fact a state initially
labeled by the quantum numbers $n_{0}$ and $S_{0}$ is expressible in
terms of the ring stationary eigenstates as
\begin{equation}
\Psi_{n_{0}}^{S_{0}}(\varphi,t)=\frac{1}{\sqrt{2\pi}}
\sum_{ns}C_{n}^{S}(n_{0},S_{0},t)e^{i(n+1/2)\varphi}e^{-iE_{n}^{S}t/\hbar}|\nu^{S}\rangle,
\label{wavef1}
\end{equation}
where $|\nu^{S}\rangle=\nu^{S}(\gamma,\varphi)$. From Eq. (\ref{matching}) we deduce
the expansion coefficients
\begin{equation}
C_{n}^{S}=\left\{
\begin{array}{l l}
\delta_{SS_{0}}\delta_{nn_{0}}, & \text{for t}\leq0,\\
\delta_{SS_{0}}i^{n_{0}-n}J_{n_{0}-n}(\alpha), & \text{for t}>t_0,
\end{array}
\right. \label{coeff}
\end{equation}
The energy at time $t$ is then given by the relations
\begin{eqnarray}
E_{n_{0}}^{S_{0}}(t)&=&\langle\Psi_{n_{0}}^{S_{0}}(\varphi,t)|H|\Psi_{n_{0}}^{S_{0}}(\varphi,t)\rangle
=i\hbar\langle\Psi_{n_{0}}^{S_{0}}(\varphi,t)|\frac{\partial}{\partial
t}|\Psi_{n_{0}}^{S_{0}}(\varphi,t)\rangle \notag\\
&=&\sum_{ns}E_{n}^{S}|C_{n}^{S}(n_{0},S_{0},t)|^{2}. \label{energy0}
\end{eqnarray}
Substituting Eq. (\ref{eigenenergy}) in Eq. (\ref{energy0}) we find
\begin{eqnarray}\label{energy2}
E_{n_{0}}^{S_{0}}(t)&=&\frac{\hbar\omega_{0}}{2}\left\{(n_{0}-x_{S_0})^{2}+\frac{2\alpha^{2}\;
\Theta(t-t_0)-Q_{R}^{2}}{4}\right\}; \\
E_{n_{0}}^{S_{0}}(t>t_0)&=&E_{n_{0}}^{S_{0}}(t<0)+\frac{\hbar\omega_{0}}{2}\frac{\alpha^{2}}{2}.
\label{energy3}
\end{eqnarray}
Here $\Theta(t)$ is the Heaviside step function and $x_{S}$ is given
by Eq. (\ref{xs}).

\section{Dipole moment generated by the pulse under external static
magnetic field} Having identified the time-dependent spectrum and
eigenfunctions, we focus now on the charge-dynamics and the induced
polarization. To this end we inspect the charge localization
parameter, defined as \cite{matos1,matos2,moskalenko}
\begin{eqnarray}
\langle\cos\varphi\rangle_{n_{0}}^{S_{0}}(t)&=&\int_{0}^{2\pi}d\varphi|
\Psi_{n_{0}}^{S_{0}}(\varphi,t)|^{2}\cos\varphi,
\notag \\
&=&\frac{1}{2}\sum_{ns}\left\{C_{n}^{S*}C_{n-1}^{S}e^{i(E_{n}^{S}-E_{n-1}^{S})t/\hbar}
+C_{n}^{S*}C_{n+1}^{S}e^{i(E_{n}^{S}-E_{n+1}^{S})t/\hbar}\right\}
.\label{cos1}
\end{eqnarray}
From Eq. (\ref{eigenenergy}) and relation
(\ref{coeff}) for the coefficients, the localization
parameter is deduced  after some algebra to be
\begin{eqnarray}
&&\langle\cos\varphi\rangle_{n_{0}}^{S_{0}}(t)=\alpha \, h(\Omega)\, \sin
(b)\, \cos[2(n_{0}-x_{S_0})b],
\label{cos4}\\
&&h(\Omega)=J_{0}(\Omega)+J_{2}(\Omega),\, b=\omega_{0}t/2,
\mbox{ and }\Omega=\alpha\sqrt{2(1-\cos(2b))}.
\nonumber\end{eqnarray}
 Here $J_n$ is a Bessel
function with index $n$. The partial photo-induced dipole moment
$\mu_{n_{0}}^{S_{0}}(t)$ associated with the initial state with
quantum numbers $n_{0}$, $S_{0}$ and the total HCP-induced dipole
moment along the $x$ axis $\mu^{S_{0}}(t)$ for the initial spin
$S_{0}$ read
\begin{eqnarray}
\mu_{n_{0}}^{S_{0}}(t)&=&ea\langle\cos\varphi\rangle_{n_{0}}^{S_{0}}(t),
\label{dipolemoment}\\
\mu^{S_{0}}(t)&=&\sum_{E_{n_{0}}^{S_{0}}\leq
E_{F}}f(n_{0},S_0,N,t)\mu_{n_{0}}^{S_{0}}(t).
\label{dipoletatal}
\end{eqnarray}
$f$ stands for the nonequilibrium distribution function whose derivation requires the solution of the
kinetic equations.
% The
% simplest approximation to $f$ is the relaxation time approximation
% that we applied in earlier works \cite{matos1,matos2}, i.e.
% (the decay time $\tau_{rel}$ ia assumed to be spin
% independent)
%\begin{eqnarray}
% $\mu^{S_{0}}(t)=\sum_{E_{n_{0}}^{S_{0}}\leq
% E_{F}} \, e^{-t/\tau_{rel}}\, \mu_{n_{0}}^{S_{0}}(t).$
%\label{dipolet11}
%\end{eqnarray}
%
In principle we can employ our recent approach based on the density matrix \cite{andrey06} but more
detailed knowledge on the spin-dependent decay channels in confined geometry is needed. Here we inspect the
zero temperature behaviour of the induced dipole moment and the associated emission spectrum. We expect
the qualitative features of these physical quantities to persist at finite temperatures,
as we demonstrated for the case of vanishing SOI  \cite{andrey06}.

\subsection{Spectral analysis}
In the following we focus on the spectral properties. The spin-orbit
interaction breaks the energy-degeneracy of $n$ and $-n$ states. As
evident from Eqs. (\ref{xs},\ref{energy2}) the spectrum, posses  a
symmetry axis (SA) located at $x_{S}=\phi/\phi_{0}-(1-Sw)/2$ (the
global additional pulse-associated energy  and SOI-induced energy
shift do not affect this symmetry), i.e. the static magnetic flux
and the SOI act as an effective magnetic field. In Eq.
(\ref{energy2}) $n_{0}$ is an integer, hence it is  advantageous to
introduce the integer parameter $l_{S}$ as the nearest integer that
is less than $x_{S}$ and define  ($\sigma=\uparrow$ or
$\downarrow$)
\begin{equation}
\Delta_{\sigma}=x_{S}-l_S\label{delta}
\end{equation}
whose meaning is illustrated in Fig. 2(c) and 2(d). We distinguish
four cases (a), (b), (c) and (d) corresponding to
$\Delta_{\sigma}=0,1/2$, $\Delta_{\sigma}\in\, [0,1/2]$ and
$\Delta_{\sigma}\in\, [1/2,1]$, respectively (cf. Fig. 2).
$\Delta_{\sigma}=1$ is equivalent to $\Delta_{\sigma}=0$, thus
$\Delta_{\sigma}$ is periodic with changing SA and varies within the
fundamental interval $[0,1]$. Furthermore, we introduce
\begin{equation}
\bar{\Delta}_{\sigma}=|1/2-\Delta_{\sigma}|,\label{bdelta}\end{equation}
as the distance between the SA and the half integer axis. These four
cases in Fig. 2 are valid for up and spin-down states.

%%%%%%%%%%%%%%%%%%%%%%%%%%%%%%%%%%%%%%%%%%%%%%%%%%%%%%%%%%
\begin{figure}[tbph]
\centering \includegraphics[width =13 cm, height=9 cm]{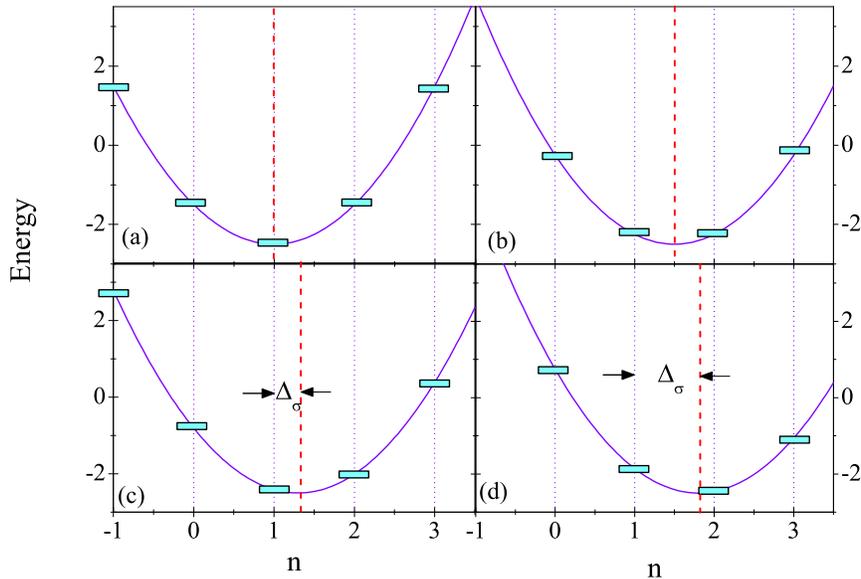}
\caption{(Color online) (a) SA has an integer value; (b) SA is a
half integer; (c) SA lies in the region $(l,l+1/2)$; (d) SA is in
the region $(l+1/2,l+1)$. The red dash lines are the SA in each
case.} \label{fig2}
\end{figure}
%%%%%%%%%%%%%%%%%%%%%%%%%%%%%%%%%%%%%%%%%%%%%%%%%%%%%%%%

\subsubsection{Spinless particles}

The two spin states, up and down, need to be considered. Allowing
each of these two spins to occupy the four configurations shown in
Fig. 2, results in 16 combinations. For simplicity, we consider at
first only one kind of spins and states which can be any of the four
cases depicted in Fig. 2. Depending on the total number of electrons
in the ring $N$ two situations are distinguished:

(1) $N$ is an even integer. For case (a) in Fig. 2 (for brevity we
refer hereafter to cases (a), (b), $\cdots$;  and write $l_S\equiv
l$, $\Delta_\sigma\equiv\Delta$, and $x_S\equiv x$) the occupied
states are at $n_{0}=\frac{1}{2}(l-m), l-(m-1), \cdots, l-1, l, l+1,
l+2, \cdots, l+(m-1), \frac{1}{2}(l+m)$. Here, e.g.
$\frac{1}{2}(l-m)$ means half occupation on the state characterized
by the quantum number $l-m$. The dipole moment in this case
$\mu_{(a)}^{e}(N,t)$ reads
\begin{eqnarray}
\mu_{(a)}^{e}(N,t)&=&\alpha\Theta(t-t_0)h(\Omega)\sin
b\sum_{n_{0}}\cos[2(n_{0}-x)b], \notag\\
&=&\alpha\Theta(t-t_0)h(\Omega)\, \sin(Nb)\, \cos (b). \label{muea}
\end{eqnarray}
 For (b), (c) and (d) cases, we have
$n_{0}= l-(m-1), \cdots, l-1, l, l+1, l+2, \cdots, l+(m-1), l+m$.
Accordingly we deduce
\begin{equation}
\mu_{(bcd)}^{e}(N,t)=\alpha\Theta(t-t_0)h(\Omega)\, \sin(Nb)\, \cos[(1-2\Delta)b],
\label{muebcd}
\end{equation}
where $\Delta=x-l$. In (a) case, $\Delta=0$, then Eq. (\ref{muebcd})
will reduce to Eq. (\ref{muea}). The dipole moment for even
occupation has in general the form
\begin{equation}
\mu^{e}(N,\Delta,t)=\alpha\Theta(t-t_0)h(\Omega)\, \sin(Nb)\, \cos[(1-2\Delta)b].
\label{mue}
\end{equation}

(2) $N$ is an odd integer. Similar steps as in the proceeding case can be
performed, leading us to conclude that
\begin{equation}
\mu^{o}(N,\bar{\Delta},t)=
\alpha\Theta(t-t_0)h(\Omega)\sin(Nb)\, \cos[(1-2\bar{\Delta}) b]. \label{muo}
\end{equation}

\subsubsection{Spin-$\frac{1}{2}$ particles and  $\phi\neq 0$}
For spin-$\frac{1}{2}$ particles the energy spectrum is
spin-dependent. The position of the spectrum SA is different for the
two spin states, i.e. $x_{\uparrow}=\phi/\phi_{0}+(w-1)/2$ and
$x_{\downarrow}=x_{\uparrow}-w$. Thus, the relative distance between
these two SAs depends on the external flux and SOI parameter $w$.
Scanning these parameters the spectrums of up and down spins are
tuned to  any of the four cases shown in Fig. 2. In what follows  we
use for cases  in Fig. 2 that correspond to the spin-down and the
spin-up states the symbols (ij) (i,j=a,b,c,d); e.g. the case (ab)
means the down-spin spectrum corresponds to the case (a) whereas the
up-spin spectrum is as in case (b); hence there are 16 combinations
of such pairs to be considered. In the following, we consider four
different occupations in the ring for up and spin-down states and
these 16 combinations in detail.

(1) \textit{Even number of pairs}. For  $N=4m$, where
$m$ is an integer  ($N$ is the total number of electrons in the
ring) there are $2m$ spin-up particles and $2m$ spin-down particles
 occupying the respective spectrums, meaning that
\begin{equation}
\mu^{\sigma}(N,t)=\mu^{e}(N/2,\Delta_{\sigma},t). \label{mueven}
\end{equation}

(2) \textit{Odd number of pairs}. If $N=4m+2$ then
 $2m+1$ spin-up particles and $2m+1$ spin-down particles populate
 the respective spectrums and
\begin{equation}
\mu^{\sigma}(N,t)=\mu^{o}(N/2,\bar{\Delta}_{\sigma},t).
\label{muodd}
\end{equation}

(3) \textit{Even number of pairs plus an extra particle}. Here we
write $N=4m+1$. i) For $2m$ spin-up particles and $2m+1$ spin-down particles we find
$\mu^{\uparrow}(2m,t)=\mu^{e}(2m,\Delta_{\uparrow},t)$, and
$\mu^{\downarrow}(2m+1,t)=\mu^{o}(2m+1,\bar{\Delta}_{\downarrow},t)$.
Cases (ab), (ac), (ad), (cb), (db) belong to this type. ii) If there
are $2m+1$ spin-up particles and $2m$ spin-down particles in the
ring we infer
$\mu^{\uparrow}(2m+1,t)=\mu^{o}(2m+1,\bar{\Delta}_{\uparrow},t)$,
and $\mu^{\downarrow}(2m,t)=\mu^{e}(2m,\Delta_{\downarrow},t)$.
Cases (ba), (ca), (da), (bc), (bd) belong to this category. iii)
If we have $2m$ spin-up particles and $2m$ spin-down particles
plus one extra particle we shall analyze the populated  state of the extra particle.
E.g., cases (aa), (bb), (cc), (dd), (cd) and (dc)are possible situations.
 Careful calculation of all  cases results in the
formula
\begin{equation}
\mu_{ex}=\alpha\Theta(t-t_0)h(\Omega)\sin(b)\cos[(2m+1-2\bar{\Delta}_{\sigma})b], \label{mu33}
\end{equation}
where $\bar{\Delta}_{\sigma}\geq\bar{\Delta}_{\bar{\sigma}}$.

(4) \textit{Odd number of pairs plus an extra particle}. For
$N=4m+3$ we distinguish: i) For $2m+1$ spin-up particles and $2m+2$
spin-down particles we infer
$\mu^{\uparrow}(2m+1,t)=\mu^{o}(2m+1,\bar{\Delta}_{\uparrow},t)$,
and $\mu^{\downarrow}(2m+2,t)=\mu^{e}(2m+2,\Delta_{\downarrow},t)$.
Cases (ab), (ac), (ad), (cb) and (db) belong to this type. ii) For
$2m+2$ spin-up and $2m+1$ spin-down particles we deduce
$\mu^{\uparrow}(2m+2,t)=\mu^{e}(2m+2,\Delta_{\uparrow},t)$, and
$\mu^{\downarrow}(2m+1,t)=\mu^{o}(2m+1,\bar{\Delta}_{\downarrow},t)$.
Cases (ba), (bc), (bd), (ca) and (da) are examples for this
situation. iii) For $2m+1$ spin-up and for $2m+1$ spin-down
particles plus an extra particle (cf.  (bb), (cc), (dd), (cd) and
(dc)) we find for the dipole moment
\begin{equation}
\mu_{ex}=\alpha\Theta(t-t_0)h(\Omega)\sin(b)
\cos[(2m+1+2\bar{\Delta}_{\sigma})b] \label{muex4}
\end{equation}
where $\bar{\Delta}_{\sigma}\leq\bar{\Delta}_{\bar{\sigma}}$.
\subsubsection{spin $\frac{1}{2}$ particles with ($\phi=0$)}
If $\phi=0$ then $x_{\uparrow}=\frac{w-1}{2}$,
$x_{\downarrow}=-\frac{1+w}{2}$, and
$\Delta_{\uparrow}+\Delta_{\downarrow}=1$ applies. The symmetry axes
for up and down-spin states have the same distance from the nearest
integer axes to the left and to the right sides to the SAs. The
dipole moments are spin degenerate.

\noindent
(1) \textit{Even number of pairs and $\phi=0$.} For $N=4m$
we find
$\mu^{\uparrow}(N,t)$=$\mu_{\uparrow}^{e}(2m,\Delta_{\uparrow},t)$.\\
(2) \textit{Odd number of pairs and $\phi=0$.} For $N=4m+2$ we
conclude
$\mu^{\uparrow}(N,t)$=$\mu_{\uparrow}^{o}(2m+1,\bar{\Delta}_{\uparrow},t)$.\\
(3) \textit{Even number of pairs plus an extra particle.} If
$N=4m+1$, only possible combinations are (aa), (bb), (cd) and (dc).
The spin degeneracy of the extra particle is caused by the crossing
of the energy levels with opposite spins. The dipole moment reads
\begin{equation}
\mu^{\uparrow}(N,t)=\frac{1}{2}\alpha\Theta(t-t_0)h(\Omega)
\left\{\sin(2mb)\cos[(1-2\Delta_{\uparrow})b]+\sin[(2m+1)b]
\cos(2\Delta_{\uparrow}b)\right\}. \label{mu3phi0}
\end{equation}
If $\Delta_{\uparrow}=0$, Eq. (\ref{mu3phi0}) delivers the dipole
moment for the case (aa) whereas  $\Delta_{\uparrow}=1/2$ applies
for the case (bb) and others for case (cd) and (dc).\\
(4) \textit{Odd number of pairs plus an extra particle.} For
$N=4m+3$, only (aa), (bb), (cd) and (dc) are applicable. The dipole
moment is
\begin{equation}
\mu^{\uparrow}(N,t)=\frac{1}{2}\alpha\Theta(t-t_0)h(\Omega)
\left\{\sin[(2m+1)b]\cos[(1-2\bar{\Delta}_{\uparrow})b]+\sin[(2m+2)b]\cos
(2\bar{\Delta}_{\uparrow}b)\right\}. \label{mu4phi0}
\end{equation}
In the event $\bar{\Delta}_{\uparrow}=0$, Eq. (\ref{mu4phi0}) yields
the dipole moment for case (bb) whereas
$\bar{\Delta}_{\uparrow}=1/2$ is valid for the case (aa) and others
for case (cd) and (dc).

\section{Numerical Results and Discussions}
\subsection{Experimental feasibility and general remarks}
In this section we present and analyze   numerical results for the
HCP-induced polarization of a ballistic quantum ring. In view of an
experimental realization it is important to identify the realistic
range of the parameters such as the strength of the SOI and the
associated quantities. The range of the ring size and external field
strength are chosen according to current experimental feasibility.

The Rashba SOI was already investigated for numerous  semiconductor
quantum wells such as  In$_{x}$Ga$_{1-x}$As/InP quantum wells
\cite{thsch}, In$_{0.53}$Ga$_{0.47}$As/In$_{0.52}$Ga$_{0.48}$As
heterostructures \cite{ingaas}, and GaSb/InAs/GaSb quantum wells
\cite{luo}. Spin-interference effects in a ring with Rashba SOI
which was built in a InGaAs/InAlAs heterostructure was the subject
of Ref. [\onlinecite{nitta03}], and the Aharonov-Casher phase was
inspected in II-VI semiconductor quantum rings, such as HgTe/HgCdTe
ring \cite{konig}. In the context of the present work it is
important to estimate the realistic range for the spin orbit angle
$\gamma$ for the relevant semiconductor materials: For
In$_{x}$Ga$_{1-x}$As/InP quantum well \cite{thsch}, $\alpha_{R}$
varies in the range $[0.7\times10^{-11}eV\cdot m,
1.1\times10^{-11}eV\cdot m]$ when an applied gate voltage varies
from +1.5 V to -2.5 V; this corresponds to $\gamma$ being in the
range $[-34^{\circ}, -47^{\circ}]$, $[-74^{\circ}, -79^{\circ}]$,
and $[-82^{\circ}, -85^{\circ}]$ for  rings with 100 nm, 500 nm, and
1 $\mu m$ radius respectively ($m^{*}=0.037m_{0}$, $m_0$ is the free
carrier mass). For In$_{0.53}$Ga$_{0.47}$As/In$_{0.52}$Ga$_{0.48}$As
materials \cite{ingaas}, $\alpha_{R}$ varies in
$[0.5\times10^{-11}eV\cdot m, 1.0\times10^{-11}eV\cdot m]$ if the
external gate voltage changes from +1.5 V to -1 V; correspondingly
$\gamma$ varies in $[-33^{\circ}, -52^{\circ}]$, $[-73^{\circ},
-81^{\circ}]$, and $[-81^{\circ}, -86^{\circ}]$ for
 rings with 100 nm, 500 nm, and 1 $\mu m$ radius respectively
($m^{*}=0.05m_{0}$). For GaSb/InAs/GaSb quantum well \cite{luo} the
values of $\gamma$ are on the same order as above.

With regard to available pulses, a wide range of pulse durations and
strengths have been realized
\cite{hcp,seqhcp1,seqhcp2,seqhcp3,seqhcp4}. As clear from the above
analysis the pulse parameter which is decisive for the electron
dynamics is $\alpha$ as given by Eq. (\ref{alpha}). For pulse with a
sin-square shape and  a duration of $\tau_d=1$~ps we achieve for a
ring with radius $1~\mu$m a transferred action of $\alpha_1=0.1$,
$\alpha_2=1$, and $\alpha_3=10$ if the peak electric field strength
is tuned to respectively  $E_1=1.32~$V/cm, $E_2=13.2~$V/cm, and
$E_3=132~$V/cm.  It is this range of $\alpha$ which we use in the
present numerical calculations. In the figures below we just provide
$\alpha$.

From a general point of view we can expect four frequency scales to be relevant for the time-dependent charge and spin dynamics:\\
1). The global energy scale is set by the size of the system. Hence, the fundamental frequency is  given by
$\omega_0=\hbar/(m^*a^2)$ and the natural time scale is  $t_p=4\pi/\omega_0$ which for the ring
sizes at hand is tens of picoseconds.\\
2). As for any fermionic system, the Fermi energy $E_F$ sets the scale for the fast (femtosecond) charge dynamics
    associated with excitation near $E_F$.
    Since we are dealing with an isolated ring the Fermi energy is expressible in terms of the number of particles $N$ and the relevant frequency is therefore $N \omega_0$.\\
3). A Further frequency $\omega_R$ is associated with
SOI-influenced dynamics; $\omega_R=2\alpha_R/(\hbar a)$.
     For our systems $\omega_R$ can be on the order of $\omega_0$ opening the possibility
      for controlling quantum interferences, e.g. by tuning the strength of SOI (via a gate voltage).\\
4). Further modifications are brought about by the applied pulse
which (for the pulses
     of interest here) induces a multitude of excitations near $E_F$  with further associated frequencies, as detailed below.

These physical expectations are confirmed by the general structure
of the calculations for the induced dipole moments according to  Eqs.
(\ref{cos4},\ref{muea},\ref{mue},\ref{muo},\ref{mu33},\ref{muex4})

\subsection{Zero static magnetic field}
At first we investigate the case of vanishing static flux
($\phi=0$). In Fig. 3 the time-dependence of the dipole moment  is
shown for different spin-orbit angles $\gamma$ and  pulse strengths
$\alpha$. Because of the spin-degeneracy we consider only one spin
channel. The  total number of particles is $N=4m$. The  time is
measured in units of the system's time scale   $t_{p}=4\pi
m^{*}a^{2}/\hbar$.
%%%%%%%%%%%%%%%%%%%%%%%%%%%%%%%%%%%%%%%%%%%%%%%%%%%%%%%%%%
\begin{figure}[tbph]
\centering \includegraphics[width =14 cm, height=10 cm]{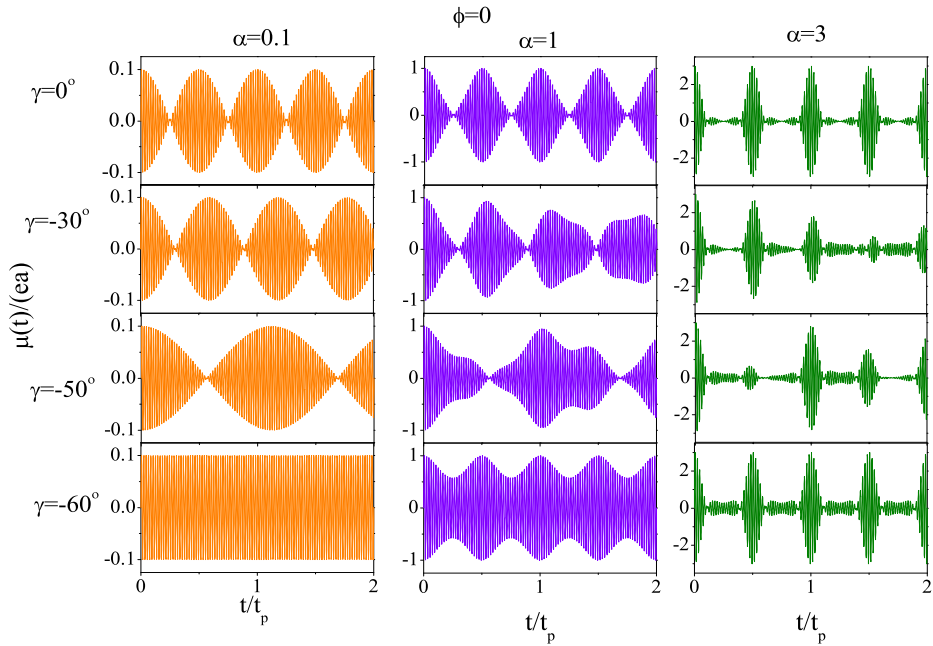}
\caption{(color online) The time-dependence (in units of
$t_p=4\pi/\omega_0$) of the net pulse-induced dipole moment $\mu(t)$
(in units of $ea$) in the up or down-spin channel at different
$\gamma$ and $\alpha$ as depicted on the figure. The total number of
particles is $N=100$ and $\phi=0$.} \label{fig3}
\end{figure}
%%%%%%%%%%%%%%%%%%%%%%%%%%%%%%%%%%%%%%%%%%%%%%%%%%%%%%%%
%%%%%%%%%%%%%%%%%%%%%%%%%%%%%%%%%%%%%%%%%%%%%%%%%%%%%%%%%%
\begin{figure}[tbph]
\centering \includegraphics[width =15 cm, height=9 cm]{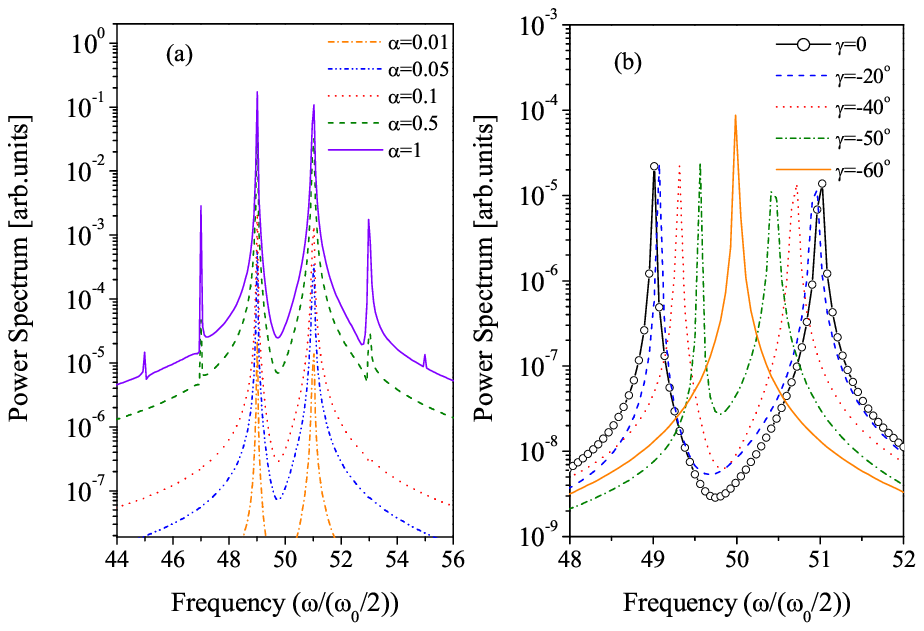}
\caption{(color online) The emission spectrum for different pulse
strengths (quantified by $\alpha$) and SOI strengths (indicated by
$\gamma$). In (a) $\alpha$ is varied at $\gamma=0$ whereas in (b)
$\gamma$ is variable at fixed $\alpha=0.01$; in both cases $N=100$.
} \label{ep}
\end{figure}
%%%%%%%%%%%%%%%%%%%%%%%%%%%%%%%%%%%%%%%%%%%%%%%%%%%%%%%%

The four time scales mentioned above are clearly visible. As
inferred from Eqs. (\ref{muea}, \ref{mue}, \ref{muo}, \ref{mu33},
\ref{muex4}) the net time-dependent dipole moment grows linearly
with the pulse field strength at small $\alpha$ (note for
$\alpha\ll1/2$,  $\Omega\ll1$, we have $h(\Omega)\approx1$). The
fast oscillation in the dipole moment are related the transition
between the levels near $E_F$ (Rabi flopping). With increasing
$\alpha$ more levels are excited giving rise to higher contributing
harmonics (cf. $\alpha=1$ and $\alpha=3$ in Fig. 3 and also Fig. 4).

The SOI strength (quantified by $\gamma$) has a dramatic influence
on the low frequency modulation of the dipole-moment envelope. The
low frequency is associated with the difference between the
frequencies of the involved levels near $E_F$, which is  on the
scale of $\omega_0$. In fact, as inferred from Eq. (\ref{mue})
$\gamma$ can be tuned as to influence the phases of the involved
wave functions changing thus the interference pattern and removing
eventually the slow oscillations altogether. This happens at
$\gamma=-60^\circ$ as shown in Fig. 3. When  $|\gamma|$ is increased
further the slow modes appear again. This is insofar important as
$\gamma$ can be modified by an external gate voltage offering thus
the possibility of engineering the emission spectrum via an applied
static electric field and opening the way for testing experimentally
our theoretical predictions. For this reason  we inspect the power
spectrum \cite{alex_power} produced by the non-equilibrium charge
oscillations in the QR by evaluating
\begin{equation}
P(\omega)=|\mu(\omega)|^{2}, \quad \mu(\omega)=\int_{-\infty}^{\infty}\mu(t)e^{-i\omega t}dt
.\label{ps}\end{equation}
Fig. 4 shows the power spectrum for different strengths of the pulse
$\alpha$. As evident from these calculations the frequency scale is
set by $\omega_0/2$. In Fig. 4 the number of particles occupying the
single spin states is N=50, for small $\alpha$ only the state at
$E_F$ is excited leading to the appearance of two frequencies at
$(N-1)\omega_0/2$ and  $(N+1)\omega_0/2$. With increasing $\alpha$,
more  levels are excited and correspondingly further frequencies in
unit of $\omega_0/2$ emerge at $47, 45, \cdots$ and $53, 55,
57,\cdots$ (see Fig. 4(a)). In this context we note that a very
short HCP contains almost all frequencies. Nevertheless, at low
intensities only a limited number of states in the ring can be
excited. The reason is obvious from Eq. (\ref{energy3}). The highest
energy level achieved upon excitation is
$E_F+\hbar\omega_0\alpha^2/4$. Hence there is an excited energy
cut-off set by the field intensity (available photons) and limits
consequently for a certain $\alpha$ the number of possible
frequencies  observable in the power spectrum (as seen in Fig. 4).

Fig. 4(b) shows how the SOI shifts of the frequencies: with
increasing $|\gamma|$ the frequency at 49 moves towards a higher
frequency while the frequency at 51 moves to a lower frequency. For
the angle $\gamma=-60^{\circ}$ we infer
 $\Delta=1/2$  and the two
frequencies merge into one frequency. Further increasing  the
SOI strength, i.e. $\Delta>1/2$, the frequency peak from 51 continues moving
to 49, while the frequency peak from 49  approaches 51 until they coincide for
 $\Delta=1$
(or $\Delta=0$). This behavior
 of frequencies is  repeated periodically with increasing SOI.

%%%%%%%%%%%%%%%%%%%%%%%%%%%%%%%%%%%%%%%%%%%%%%%%%%%%%%%%%%
\begin{figure}[tbph]
\centering \includegraphics[width =9 cm, height=9 cm]{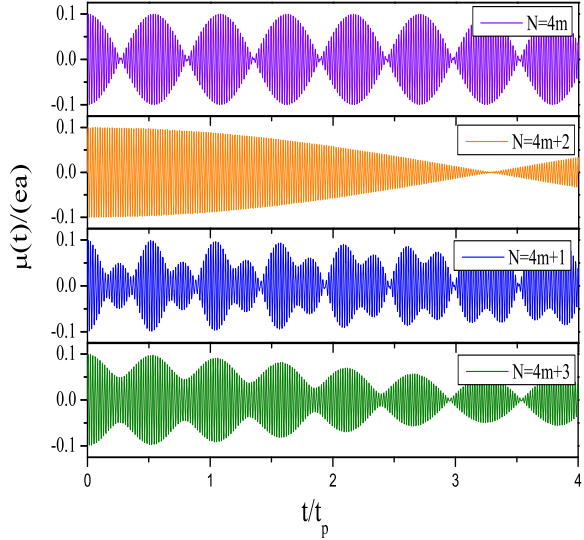}
\caption{(color online) The time-dependence of dipole moment is
shown at different occupation numbers. Here $\gamma=-70^{\circ}$,
$\alpha=0.1$ and $m=25$.  Units are as in Fig. 3.}
\label{dtphi0n234}
\end{figure}
%%%%%%%%%%%%%%%%%%%%%%%%%%%%%%%%%%%%%%%%%%%%%%%%%%%%%%%%

As  detailed above the  dipole moment depends sensitively on the
occupation numbers. Fig. 5  shows an example of the dipole-moment
dynamics  for $N=4m+2$  contrasted with the case $N=4m$. We remark
that since $\Delta+\bar\Delta=1/2$ we have $\Delta=0(\frac{1}{2})$
for $\bar\Delta=\frac{1}{2}(0)$. Thus for $N=4m$,  when the dipole
moment increases with $\Delta$ in $\Delta\in[0,\frac{1}{2}]$ the
dipole moment  decreases  for the case  $N=4m+2$. Furthermore, the
parity of the occupation number is very important for the property
of dipole moment of the ring. E.g.,
 for $\Delta\approx\bar\Delta$, the case  $N=4m$
($N=4m+1$) behaves  similarly  to $N=4m+2$ ($N=4m+3$) in which cases
the occupations are even (odd). However, the even case is
qualitatively different from the odd case (not shown in Fig. 5). If
$\Delta\neq\bar\Delta$ the oscillations of the dipole moment are
different in all the cases depicted in Fig. 5.

\subsection{Finite static magnetic field}
%%%%%%%%%%%%%%%%%%%%%%%%%%%%%%%%%%%%%%%%%%%%%%%%%%%%%%%%%%
\begin{figure}[tbph]
\centering \includegraphics[width =10 cm, height=6 cm]{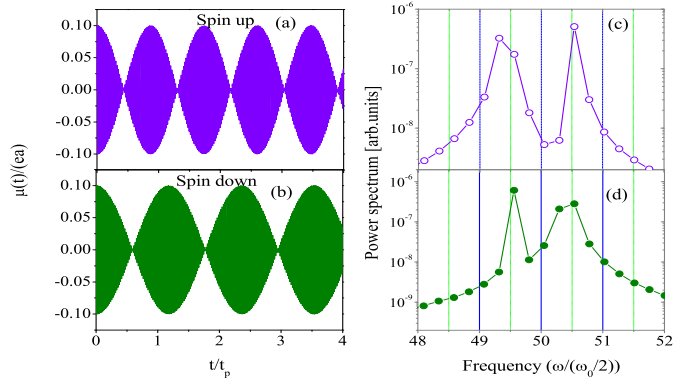}
\caption{(color online) The time-dependence of the dipole moment for
up and down spins under a nonzero static magnetic field are shown in
(a) and (b). The power spectra corresponding to the up and down
spins are shown in (c) and (d) respectively. Here $\phi=0.25$,
$\gamma=-70^{\circ}$, $\alpha=0.1$ and $N=100$. } \label{dtphin0n1}
\end{figure}
%%%%%%%%%%%%%%%%%%%%%%%%%%%%%%%%%%%%%%%%%%%%%%%%%%%%%%%%
A static external magnetic field  lifts the spin degeneracy of the
dipole moment. The induced dipole moment dynamics becomes spin
dependent (cf. Fig. 6). The slow-frequency shift between the
dipole-moment oscillations in the up-spin and down-spin channels is
readily understood from the energy splitting (cf. Eq.
(\ref{energy3})). The power spectrums for the up and down-spins
dipole  oscillations (Fig. 6(c) and 6(d)) reveals clearly the
frequency shift.

For more insight into the role of the SOI
we investigate
the spin-resolved local charge density   before and after the pulse. The probability density associated with
the level labeled by $n_{0}$ and $S_{0}$ is
$\rho_{n_{0}}^{S_{0}}(\varphi,t)=|\Psi_{n_{0}}^{S_{0}}(\varphi,t)|^{2}$.
Before the pulse the charge density  is a unit charge uniformly distributed around the ring, i.e. $\rho_{0}=1/(2\pi)$.
 Upon applying the pulse we find
\begin{equation}
\rho_{n_{0}}^{S_{0}}(\varphi,t>0)=\rho_{0}+\vartriangle\rho_{n_{0}}^{S_{0}}(\varphi,t>0),
\label{densityall}
\end{equation}
where the second term is the spin-resolved, field-induced charge
density variation (ICDV) which we evaluated and find
\begin{eqnarray}
\vartriangle\rho_{n_{0}}^{S_{0}}(\varphi,t>0)&=&\frac{1}{\pi}\{\sum_{n=0}^{\infty}J_{2n+1}[2\alpha\sin(2n+1)b]\cos(2n+1)[\varphi-2(n_{0}-x_{S_{0}})b]
\notag\\
&&+\sum_{n=1}^{\infty}J_{2n}[2\alpha\sin(2nb)]\cos2n[\varphi-2(n_{0}-x_{S_{0}})b]\}.
\label{vdensity}
\end{eqnarray}
The total spin-resolved charge density is obtained from a sum over
the occupied levels $
\rho^{S_{0}}(\varphi,t>0)=N_{S_{0}}\rho_{0}+\vartriangle\rho^{S_{0}}(\varphi,t>0),$
where $N_{S_{0}}$ is the total particle number in  the  $S_{0}$ spin
channel. The term $\vartriangle\rho^{S_{0}}(\varphi,t>0)$ for the even
pair occupation case is
\begin{eqnarray}
\vartriangle\rho^{S_{0}}(\varphi,t>0)&=&\frac{\alpha}{\pi}\{\sum_{n=0}^{\infty}\frac{1}{2n+1}[J_{2n}(\Omega_{1n})+J_{2n+2}(\Omega_{1n})]
\sin[N_{S_{0}}(2n+1)b]\cos(2n+1)\varphi_{S_{0}}(t)
\notag\\
&&+\sum_{n=1}^{\infty}\frac{1}{2n}[J_{2n-1}(\Omega_{2n})+J_{2n+1}(\Omega_{2n})]\sin(N_{S_{0}}2nb)\cos2n\varphi_{S_{0}}(t)\},
\label{vdensityall}
\end{eqnarray}
where $\Omega_{1n}=a\alpha\sin(2n+1)b$,
$\Omega_{2n}=a\alpha\sin(2nb)$, and
$\varphi_{S_{0}}(t)=\varphi-(1-2\Delta_{S_{0}})b$.
%%%%%%%%%%%%%%%%%%%%%%%%%%%%%%%%%%%%%%%%%%%%%%%%%%%%%%%%%%
\begin{figure}[tbph]
\centering \includegraphics[width =8.5 cm, height=6 cm]{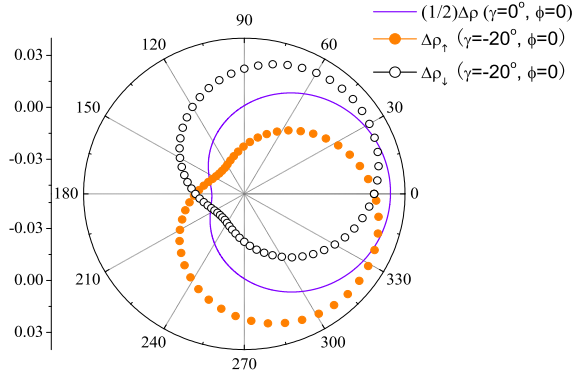}
\caption{(color online) A snap-shot of the pulse-induced charge
density variation (ICDV) as a function of  the azimuthal angle
$\varphi$ at the time $t/t_{p}=2.005$. The pulse parameter is
$\alpha=0.1$, and $N=100$.  Full (open) dots show the ICDV in the up
(down) spin channel. Solid curve stands for the spin-averaged ICDV.}
\label{densityangle}
\end{figure}
%%%%%%%%%%%%%%%%%%%%%%%%%%%%%%%%%%%%%%%%%%%%%%%%%%%%%%%%

Eq. (\ref{vdensityall}) indicates a spin-dependent phase of
$\vartriangle\rho^{S_{0}}$ caused by the  interplay of  SOI, the
static magnetic field, and the pulse field. Interestingly the phase
shift evolves with time. Fig. 7 shows the time-evolution of the
spin-dependent ICDV. In all cases depicted in Fig. 7 the static flux
is absent. For $\gamma=0$ we observe, as expected, how the pulse
kicks the charge density along the pulse-polarization axis.  For the
given time moment, the missing charge density around $\varphi=\pi$
is pushed to the region around $\varphi=0$. For a finite  spin orbit
interaction ($|\gamma|>0$) we observe a spin splitting of ICDV,
meaning that the pulse induces temporally and locally a finite spin
polarization $P=\rho_{\uparrow}-\rho_{\downarrow}$, even though the
system is initially paramagnetic. The time integral of the
pulse-induced, local spin polarization vanishes however.
Technically, we infer that the SOI results in  a SU(2) flux that
produces opposite phase shifts on the azimuthal angle with the same
magnitude for a zero-static magnetic field, i.e. $\varphi_{\uparrow
(\downarrow)}=\varphi\mp(1-2\Delta_{\uparrow})b$. These shifts cause
a rotation  around the ring  of the symmetry axes of the spin-up and
the spin-down densities respectively clockwise and anticlockwise
(see Fig. 7). When $\gamma=-60^{\circ}$ ($\Delta_{\uparrow}=1/2$),
the up and down ICDV merge into one curve after one period.  The
periodic rotation is subject to  the condition
$\Delta_{\uparrow}=1/2$. The spin-averaged ICDV, i.e.
$\vartriangle\rho=\vartriangle\rho_{\uparrow}+\vartriangle\rho_{\downarrow}$,
is always symmetric with respect to $x$ axis. At $\varphi=0$ and
$\pi$ ICDV is spin-degenerate for all times ($P(t)=0$).

The time evolution of the spin-resolved ICDV is depicted in Fig. 8.
The symmetry axes of the pulse-induced spin-up and spin-down ICDV
rotate  around the ring respectively anticlockwise and clockwise
with time (which is the opposite behaviour when increasing SOI).
Accordingly the total ICDV oscillates along the $x$ axis and
possesses a left-right symmetry with respect  $x$. The local and
temporal spin polarization $P$ is symmetric to $y$ axis and
oscillates along it with the same frequency as $\rho_{\uparrow}$ (or
$\rho_{\uparrow}$).
%%%%%%%%%%%%%%%%%%%%%%%%%%%%%%%%%%%%%%%%%%%%%%%%%%%%%%%%%%
\begin{figure}[tbph]
\centering \includegraphics[width =10 cm, height=8 cm]{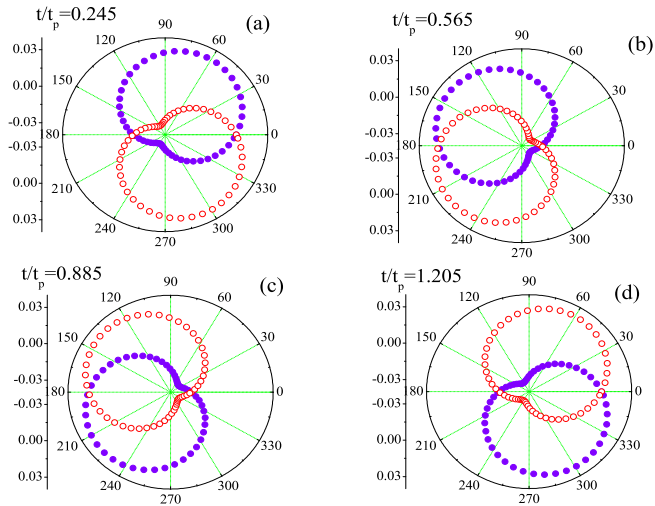}
\caption{(color online) The time and spatial dependence of the ICDV
shown in polar coordinates in (a), (b), (c) and (d) at different
times after the pulse. Filled circles represent the up spin ICDV and
the open circles indicate the down spin ICDV. The parameters are
$\alpha=0.1$, $\phi=0$, $\gamma=-40^{\circ}$ and $N=100$.}
\label{densityangletime}
\end{figure}
%%%%%%%%%%%%%%%%%%%%%%%%%%%%%%%%%%%%%%%%%%%%%%%%%%%%%%%%

\section{Conclusios}
In summary, we investigated the dynamical response of a quantum ring
with spin-orbit interaction upon the application of a linearly
polarized time-asymmetric weak electromagnetic pulse. It is found
that the dipole moment along the pulse polarization axis is spin
degenerate when there is no external magnetic field. The dipole
moment oscillates with time and the SOI provides an envelope
function or shifts of the oscillation  frequencies of dipole moment.
Stronger pulse fields can excite higher and lower harmonics. And the
envelope functions for different parity of the occupation number on
the ring are different. When a static external magnetic field is
applied, the spin degeneracy is removed. The spatial and temporal
dependence of the pulse induced charge-density variation (ICDV)
indicates that the SOI results in a SU(2) flux leading to a
splitting of the phases of the up and down spins states. The
symmetric axes of the ICDV for the up and down spins are rotated
equally in clockwise and anticlockwise when increasing SOI. The
total ICDV and the local and temporal polarization of the charge
density are symmetric to  respectively the light polarization axis
and the axis perpendicular. The pulse-induced polarization is
experimentally accessible by measuring the power spectrum of the
emitted radiation.

We thank A.S. Moskalenko and A. Matos-Abiague for fruitful discussions.
The work is support by the cluster of excellence "Nanostructured  Materials"
of the state Saxony-Anhalt.
%\newpage


\begin{thebibliography}{}
%
%
\bibitem{heizel} T. Heinzel, \textit{Mesoscopic Electronics in Solid State
Nanostructures} (Wiley-VCH Verlag,  Weinheim, 2003).
%
\bibitem{imry} Y. Imry, \textit{Introduction to Mesoscopic Physics} (Oxford University Press, Oxford,
2002).
%
%%
%%%%   Exp.  %%%%%%%%
%
%
\bibitem{exp1}
L. W. Yu,  \textit{et al.},
 % New self-limiting assembly model for Si quantum rings on Si(100)
  Phys. Rev. Lett. \textbf{98}    166102   (2007); Advanced Materials   \textbf{19}, 1577 (2007);
 %
 \bibitem{exp2} D. Mailly, C. Chapelier,  and A. Benoit,
%Experimental observation of persistent currents in GaAs-AlGaAs single loop.
Phys. Rev. Lett. {\bf 70}, 2020 (1993).
%
\bibitem{exp3} W. Rabaud, \textit{et al}., Phys. Rev. Lett. \textbf{86}, 3124 (2001).
%
%
\bibitem{exp4}
A. Fuhrer \textit{et al}.,
% S. L\"uscher, T. Ihn, T. Heinzel, K. Ensslin, W.
%Wegscheider, M. Bichler,
%Energy spectra of quantum rings
Nature \textbf{413}, 822 (2001).
%
\bibitem{exp5} A. Lorke \textit{et al.},
% Spectroscopy of nanoscopic semiconductor rings.
 Phys. Rev. Lett. \textbf{84}, 2223
 %-2226
(2000).
%
%
\bibitem{nitta99} J. Nitta, F. E. Meijer, and H. Takayanagi, Appl.
Phys. Lett. \textbf{75}, 695 (1999).
%
%
%
%%%%   Theo - nonequilibrium  %%%%%%%%
%
%
%
%
\bibitem{t1} V. E. Kravtsov, and V. I. Yudson, Phys. Rev. Lett. \textbf{70}, 210 (1993);
O. L. Chalaev, and V. E. Kravtsov, {\it ibid} \textbf{89}, 176601
(2002).
%
\bibitem{t2} O. L. Chalaev, and V. E. Kravtsov, Phys. Rev. Lett. \textbf{89},
176601 (2002).
%
%
\bibitem{t3} P. Kopietz, and A. V\"olker, Euro. Phys. J. B \textbf{3}, 397 (1998).

\bibitem{t4} M. Moskalets, and M. B\"uttiker, Phys. Rev. B \textbf{66}, 245321 (2002).
%
%
\bibitem{t5} L. I. Magarill, and A. V. Chaplik, JETP Lett. \textbf{70}, 615 (1999).
%
%
%  Newly added
\bibitem{t5a} V. Gudmundsson, C. -S. Tang, and A. Manolescu, Phys.
Rev. B \textbf{67}, 161301(R) (2003); S. S. Gylfad\'{o}ttir, et.
al., Physica Scripta \textbf{T114}, 41 (2004); Physica E
\textbf{27}, 278 (2005).
%
%
\bibitem{matos05}  A. Matos-Abiague, and J. Berakdar, Eorophys. Lett.
\textbf{69}, 277 (2005).

\bibitem{matos2} A. Matos-Abiague, and J. Berakdar, Phys. Rev. Lett.
\textbf{94}, 166801 (2005).

\bibitem{matos1} A. Matos-Abiague, and J. Berakdar, Phys. Rev. B
\textbf{70}, 195338 (2004).

%
%
\bibitem{t6} Y. V. Pershin, and C. Piermarocchi, Phys. Rev.
 B \textbf{72}, 245331 (2005).
%
 \bibitem{t7} I. Barth, J. Manz, Y. Shigeta, and K. Yagi,
 J. Am. Chem. Soc. \textbf{128}, 7043 (2006).
%
 \bibitem{andrey06} A. S. Moskalenko, A. Matos-Abiague, and
 J. Berakdar, Phys. Rev. B \textbf{74}, 161303 (2006).





\bibitem{hcp}  D. You,  R. R. Jones, P. H. Bucksbaum  and D. R. Dykaar, Opt.
Lett. \textbf{18}, 290 (1993).

\bibitem{seqhcp1}  T. J. Bensky,  G.  Haeffler, and R. R.  Jones, Phys. Rev.
Lett. \textbf{79}, 2018 (1997).

\bibitem{seqhcp2}   A. Wetzels,  A. G\"{u}rtel,  H. G. Muller, and L.
 D. Noordam, Eur. Phys. J. D \textbf{14}, 157 (2001).

\bibitem{seqhcp3} M. T. Frey \textit{et al.}, Phys. Rev. A \textbf{59}, 1434 (1999).

\bibitem{seqhcp4}  H. Maeda,  J. Nunkaew, and  T. E. Gallagher, Phys. Rev. A \textbf{75}, 053417
(2007).

%
%  Newly added
%
\bibitem{meir} Y. Meir, Y. Gefen, O. Entin-Wohlman, Phys. Rev. Lett.
\textbf{63}, 798 (1989).

\bibitem{chaplik} A. V. Chaplik, and L. I. Magarill, Superlattice and Microstructures, \textbf{18}, 321
(1995).
%
%

\bibitem{splett} J. Splettstoesser, M. Governale, and U.
Z\"{u}licke, Phys. Rev. B \textbf{68}, 165341 (2003).

\bibitem{frustaglia} D. Frustaglia, and K. Richter, Phys. Rev. B
\textbf{69}, 235310 (2004).

\bibitem{molnar} B. Moln\'{a}r, F. M. Peeters, and P.
Vasilopoulos, Phys. Rev. B \textbf{69}, 155335 (2004); \textbf{72},
75330 (2005).

\bibitem{foldi} P. F\"{o}ldi, \textit{et al}., Phys. Rev. B \textbf{71}, 33309 (2005); \textbf{73},
155325 (2006).

\bibitem{sheng} J. S. Sheng, and Kai Chang, Phys. Rev. B \textbf{74}, 235315
(2006).


\bibitem{meijer} F. E. Meijer, A. F. Morpurgo, and T. M. Klapwijk, Phys.
Rev. B \textbf{66}, 033107 (2002).

\bibitem{quantum optics} C. Gerry, and P. Knight, \textit{Introductory Quantum
Optics} (Cambridge University Press, 2005);

\bibitem{qo2} P. Lambropoulos, and D. Petrosyan,
\textit{Fundamentals of Quantum Optics and Quantum Information},
(Springer-Verlag, Berlin, Heidelberg 2007); M. O.
Scully, and M. S. Zubairy, \textit{Quantum Optics} (Cambridge
University Press, 1997).

\bibitem{rashbajpc} Yu A Bychkov, and E. I. Rashba, J. Phys. C: Solid
State Phys., \textbf{17}, 6039 (1984).

%
% Newly added
%
\bibitem{mizushima} M. Mizushima, Suppl. Prog. Theor. Phys. \textbf{40}, 207
(1967).
%
%

\bibitem{berakdar} A. Matos-Abiague, A. S. Moskalenko, and J. Berakdar,
\textit{Ultrafast dynamics of nano and mesoscopic systems driven by
asymmetric electromagnetic pulses} In: \textit{Current Topics in
Atomic, Molecular and Optical Physics} (Eds.) Sinha, C. and
Bhattacharyya, S., (World Scienctific, London,  2007).


%\bibitem{wendler} L. Wendler, and V. M. Fomin, A. A. Krokhin, Phys.
%Rev. B \textbf{50}, 4642 (1994).

\bibitem{moskalenko} A. S. Moskalenko, A. Matos-Abiague and J.
Berakdar, Euro. Phys. Lett., \textbf{78}, 57001 (2007).
%
\bibitem{thsch} Th. Sch\"{a}pers, \textit{et al}., J. Appl. Phys.
\textbf{83}, 4324 (1998).

\bibitem{ingaas} C. M. Hu, \textit{et al}., Phys. Rev. B \textbf{60},
7736 (1999); J. Nitta, \textit{et al}., Phys. Rev. Lett.
\textbf{78}, 1335 (1997).

\bibitem{luo} J. Luo, \textit{et al}., Phys. Rev. B \textbf{41},
7685 (1990).

\bibitem{nitta03} J. Nitta, and T. Koga, J. Supercond. \textbf{16},
689 (2003).

\bibitem{konig} M K\"{o}nig, \textit{et al}., Phys. Rev. Lett. \textbf{96},
76804 (2006).

%
\bibitem{alex_power}
%Emission spectrum of a mesoscopic ring driven by fast unipolar pulses
A. Matos-Abiague, and J. Berakdar, Phys. Lett. A  \textbf{330}, 113 (2004);
%   Title: Emission spectrum of an electron in a double quantum well driven by ultrashort half-cycle pulses
%Author(s): Matos-Abiague A, Berakdar J
Physica Scripta   \textbf{T118}, 241 (2005).
%
\end{thebibliography}
\end{document}